 \def\a{\alpha}\def\b{\beta}\def\g{\gamma}\def\e{\epsilon}
  \def\L{\Lambda}
 \def\s{\sigma}\def\S{\Sigma} \def\th{\theta}\def\om{\omega}

 \def\P{\Psi} \def\ph{\phi}

\def\imo{i}

\def\be{\begin{equation}}
\def\ee{\end{equation}}
\def\bea{\begin{eqnarray}}
\def\eea{\end{eqnarray}}

\def\Re#1{\mathrm{Re}(#1)}
\def\Im#1{\mathrm{Im}(#1)}

\newcommand{\p}[1]{(\ref{#1})}

\documentclass[12pt]{JHEP3}
\usepackage{graphicx,rotate}
\usepackage{longtable}
\title{High overtones of Schwarzschild-de-Sitter quasinormal spectrum}
\author{R. A. Konoplya\footnote{E-mail: konoplya\_roma@yahoo.com}~~and~A.
Zhidenko\footnote{E-mail: Zhidenko@ff.dsu.dp.ua}\\
Department of Physics, Dniepropetrovsk National University\\St.
Naukova 13, Dniepropetrovsk 49050, Ukraine}

\abstract{ We find the high overtones of gravitational and
electromagnetic quasinormal spectrum of the Schwarzschild-de
Sitter black hole. The calculations show that the real parts of
the electromagnetic modes asymptotically approach zero. The
gravitational modes show more peculiar behavior at large $n$: the
real part oscillates as a function of imaginary even for very high
overtones and these oscillations settles to some ``profile'' which
just repeats itself with further increasing of the overtone number
$n$. This lets us judge that $\Re\omega$ is not a constant as $n
\rightarrow \infty$ but rather some oscillating function. The
spacing for imaginary part $ \Im{\omega_{n+1}}-\Im{\omega_{n}}$
for electromagnetic perturbations at high $n$ slowly approach
$k_{e}$ as $n \rightarrow \infty$, where $k_{e}$ is the surface
gravity. In addition we find the lower QN modes for which the
values obtained with numerical methods are in a very good
agreement with those obtained through the  6th order WKB
technique.}


\preprint{arXiv:hep-th/0402080\\JHEP 06 (2004) 037}

\begin{document}\sloppy
\section{Introduction}

Quasinormal (QN) modes of black holes are of considerable interest
recently because of their interpretation in ADS/CFT correspondence
\cite{Horowitz1}-\cite{Starinets}, and of possibility to detect
gravitational waves from black holes \cite{kokkotas-review}.
Recently it has been observed that the quasinormal modes can play
a fundamental role in Loop Quantum Gravity (LQG) \cite{Dreyer}:
for asymptotically flat black holes it has been found that the
asymptotic value of the real part of the quasinormal frequency
(i.e. when the overtone approaches infinity) coincides with the
Barbero-Immirizi parameter, which must be fixed to predict the
Bekenstein-Hawking formula for entropy within the framework of
LQG. All this stimulated development of different approaches to
calculation of quasinormal modes \cite{qnm}. In particular, the
analytical expression for asymptotically high QN modes of
D-dimensional Schwarzschild black hole was obtained in \cite{Motl
2}, \cite{Motl}.

The asymptotic overtone quasinormal behavior  was studied by
Nollert \cite{Nollert} for the Schwarzschild black hole (see [5]),
and in \cite{vitorromanjose} for the Schwarzschild anti-de Sitter
black hole. Yet there is no such study for Schwarzschild-de Sitter
black hole (SdS BH), except for near extremal case \cite{yoshida}.
By using the 6th order WKB technique the low lying modes of SdS BH
were estimated in \cite{otsuki}. In the work \cite{Moss-Norman},
the low lying QN modes for gravitational perturbations of
Schwarzschild-de Sitter black hole were obtained with the help of
the Leaver method \cite{Leaver}. In \cite{konoplya-prd5} it was
found the low lying QN modes for higher ($> 4$) dimensional
Schwarzschild black hole with different values of lambda term;
this includes cases of Schwarzschild, Schwarzschild-de Sitter, and
Schwarzschild-anti-de Sitter black holes.

For the higher QN modes of  the near extremal SdS BH, in
\cite{yoshida} it  was shown that in contrast to asymptotically
flat SBH, the QN spectrum of SdS BH have oscillatory behavior: the
real part oscillates as a function of imaginary. Recently there
appeared a lot of works using different numerical and analytical
approaches devoted to near extremal Schwarzschild-de-Sitter
quasinormal spectrum \cite{kucha mala}.

In this paper we make numerical investigation of the high
overtones of Schwarzschild-de-Sitter quasinormal spectrum. We show
that at very high overtones the real parts of the gravitational QN
modes oscillates as a function of imaginary part.  Thus the real
part of $\omega$ does not go to any constant limit and, thereby,
the interpretation of the asymptotic QN frequency as those
connected with  the Barbero-Immirizi parameter in LQG is
impossible. We also show that the real part of the electromagnetic
QN modes approaches zero as  an overtone number goes to infinity.
At high overtones the QN spectrum shows rather peculiar behavior
of some periodic weaving. The numerical results at high overtones
are in a very good agreement with recent analytical (algebraic)
equations which govern the asymptotic behavior, while the lower
overtones are in a good agreement with those obtained earlier with
the 6-th order WKB method.

The paper is organized as follows. In Sec. II there are basic
formulas of SdS background for Nollert technique. Sec. III is
devoted to the low (first ten) overtones. Sec. IV deals with very
high overtones which let us judge about overtone asymptotic
behavior.

\section{Basic equations}
The Schwarzschild-de-Sitter black hole is described by the metric
\be\label{SdS-metric}%
ds^2 = f(r)dt^2 - \frac{dr^2}{f(r)} - r^2 d\s^2; \qquad f(r) = 1 -
\frac{2M}{r} - \L\frac{r^2}{3}, \quad d\s^2 = d\th^2 + \sin^2\th
d\ph^2,
\ee%
where $M$ is the black hole mass, $\L$ is the cosmological
constant.

It is well known that the perturbation equations can be reduced to
the Schr\"{o}dinger wave-like equation
\be\label{Wave-like-equation}%
\left(\frac{d^2}{dr^{*2}} + \om^2 - V(r^*)\right)\P(r^*) = 0, \ee%
by using the so-called tortoise coordinate:
\be\label{tortoise-coordinate} dr^* = \frac{dr}{f(r)}. \ee%

Under the choice of the positive sign of the real part of
$\omega$, QNMs satisfy the following boundary conditions
\be\label{bounds}
\P(r^*) \sim C_\pm \exp(\pm\imo\om r^*), \qquad r\longrightarrow
\pm\infty, \ee corresponding to purely in-going waves at the event
horizon and purely out-going waves at the cosmological horizon.

The effective potential is given by  \be\label{potential} V(r) =
f(r)\left(\frac{l(l+1)}{r^2} - \frac{2M\e}{r^3}\right), \ee where
$\e=3$ for gravitational perturbations and $\e=0$ for
electromagnetic ones.

The appropriate Frobenius series are \be\label{Frobenius} \P(r^*)=
\left(\frac 1 r - \frac 1 {r_e}\right)^{\rho_e}\left(\frac 1
{r_c}-\frac 1 r\right)^{-\rho_c}\left(\frac 1 r+\frac 1
{r_e+r_c}\right)^{\rho_c+\rho_e}\sum_{n\geq0}a_n\left(\frac{\frac
1 r-\frac1{r_e}}{\frac1{r_c}-\frac1{r_e}}\right)^n, \ee where
$r_e$ is the event horizon, $r_c$ is the cosmological horizon,
$\rho_e$ and $\rho_c$ are determined by
\be\label{rho}%
 e^{\imo\om
r^*}=\left(\frac 1 r - \frac 1 {r_e}\right)^{-\rho_e}\left(\frac 1
{r_c}-\frac 1 r\right)^{-\rho_c}\left(\frac 1 r+\frac 1
{r_e+r_c}\right)^{\rho_c+\rho_e},
\ee%
and one can find
$$\rho_e=\frac{\imo\om}{\displaystyle 2M\left(\frac{1}{r_c}-\frac{1}{r_e}\right)\left(\frac{1}{r_c+r_e}+\frac{1}{r_e}\right)};\qquad\rho_c=\frac{-\imo\om}{\displaystyle 2M\left(\frac{1}{r_c}-\frac{1}{r_e}\right)\left(\frac{1}{r_c+r_e}+\frac{1}{r_c}\right)}.$$

Substituting \p{Frobenius} into \p{Wave-like-equation}, we obtain
the three-terms recurrent relation for $a_n$ \be\label{reccur}
a_{n+1}\a_n+a_n\b_n+a_{n-1}\g_n=0, \qquad n\geq0, \quad \g_0=0,
\ee where the coefficients $\alpha, \beta, \gamma$ have the form:
\bea\label{coefficients} \a_n&=&\frac{r_c(r_c + 2 r_e) (1 + n + 2
\rho_e)^2} {r_c^2 + r_c r_e + r_e^2} + \frac{2 r_c r_e (1 + n +
2\rho_e) }{r_c-r_e}\imo\om\\\nonumber
\b_n&=&-\frac{(n+2\rho_e)(n+2\rho_e+1)(2 r_c^2 + 2 r_c r_e  -
r_e^2)}{r_c^2 + r_c r_e + r_e^2} -l(l+1) +
\frac{r_c(r_c+r_e)}{r_c^2 + r_c r_e + r_e^2}\e
\\\nonumber\g_n&=&\frac{r_c^2 -r_e^2 }{r_c^2 + r_c r_e  + r_e^2}
((n+2\rho_e)^2-1-\e).
\eea

Following Leaver \cite{Leaver} we are searching QNMs as the most
stable roots of \be\label{continued_fraction}
\b_n-\frac{\a_{n-1}\g_{n}}{\b_{n-1}
-\frac{\a_{n-2}\g_{n-1}}{\b_{n-2}-\a_{n-3}\g_{n-2}/\ldots}}=
\frac{\a_n\g_{n+1}}{\b_{n+1}-\frac{\a_{n+1}\g_{n+2}}{\b_{n+2}-\a_{n+2}\g_{n+3}/\ldots}}.
\ee The infinite continued fraction on the right side of the
equation \p{continued_fraction} converges worse if the imaginary
part of $\om$ increases with respect to the real part. This
problem was circumvented by Nollert \cite{Nollert}. He considered
\be\label{NollertR} R_N =
\frac{\g_N}{\b_N-\frac{\a_N\g_{N+1}}{\b_{N+1}-\a_{N+1}\g_{N+2}/\ldots}}
= \frac{\g_N}{\b_N-\a_NR_{N+1}}. \ee Making use of the recurrence
relation \p{NollertR} one can find for large N:
\be\label{NollertSeries} R_N =
C_0+C_1N^{-1/2}+C_2N^{-1}+C_3N^{-3/2}\ldots \ee where
$$C_0=\frac{r_c^2-r_e^2}{r_c(r_c+2r_e)},$$
$$C_1=\pm\sqrt{\frac{2r_c^2r_e+5r_cr_e^2+2r_e^3-2(r_c^3+r_e^3)
r_e\imo\om-4(r_c+r_e)r_e^2r_c\imo\om}{r_c^3+4r_cr_e(r_c+r_e)}},
\qquad \Re C_1>0,$$ etc.

The series \p{NollertSeries} converge for $|\om|/N< A<\infty$, so
we can use this approximation for $R_N$ inside the continued
fraction for some $N\gg -\Im\om\sim n$. In practice to find an
appropriate $N$ we increase it until the result of the continued
fraction calculations does not change. If one is limited by the
near extremal SdS black hole the imaginary part, being
proportional ro surface gravity, is still small in this limit,
and, one can use the Frobenius method without Nollert modification
\cite{yoshida} which includes expansion in $N$. For non extremal
values of $\Lambda$ the situation is more complex and we have to
deal with Nollert technique as described above.

\section{Lower overtones}

Here we present results of calculation for first ten overtones for
different values of $\Lambda$ and $l$ (see Appendix in this
paper). It turned out that the lower overtones obtained here with
the help of Leaver method are in a very good agreement with those
obtained through the 6th order WKB method \cite{Zhidenko}. Thus
for example for $\Lambda =0.02$ and $l=2$ gravitational
perturbations from the 6th order WKB approach  we have for the
fundamental overtone ($n=0$) $\omega$ = $0.3384-0.817 i$ while
from the continued fraction we obtain $\omega$ =
$0.33839143-0.08175645 i$. For electromagnetic perturbations with
$l=1$ $n=0$ we have $\omega$ = $0.2259-0.0842 i$ and $\omega$ =
$0.22594346-0.08410383 i$ from the 6th order WKB formula and from
continued fractions respectively. Note that the pure imaginary
algebraically special value for $l=2$ gravitational perturbations
which corresponds to the 8th mode ``move'' to the 9th mode for
$\Lambda =0.02$ and to the higher mode for greater $\Lambda$. When
the $\Lambda$ is growing both the real and the imaginary parts of
$\omega$ are decreasing, i.e. modes damp more slowly and
oscillates with greater real frequency. In the near extremal
regime, i.e. when the $\Lambda$ term  is close to its extremal
value $1/9$ ($M=1$), the effective potential approaches the
P\"{o}schl-Teller potential and only several first modes are well
described by the formula: \be\label{ese} \om b = -\left(n+\frac 1
2\right)\imo + \sqrt{l(l+1)-\frac 1 4} \ee for scalar and
electromagnetic perturbations, and by \be\label{eg} \om b =
-\left(n+\frac 1 2\right)\imo + \sqrt{(l+2)(l-1)-\frac 1 4} \ee
for gravitational (axial) perturbations. Here
$$b = \frac{54M^3}{(r_c-3M)(r_c + 6M)}, \qquad r_c\rightarrow
3M.$$

Note that the gravitational perturbations can be divided into the
two kinds which can be treated separately: axial (symmetric with
respect to the change $\varphi \rightarrow -\varphi$) and polar.
In \cite{Zhidenko} it was shown both numerically and analytically
that there is the isospectrality between these two kinds of
perturbations, i.e. they both induce the same QN spectrum. That is
why we treat here only the axial type of gravitational
perturbations.

\section{High overtones}

Finding of very high overtones is a time consuming procedure since
the ``length'' of the continued fractions must be large enough.
When we give for example the 100000th overtone that does not mean
that we found all the previous 99999 modes; that would require an
enormous amount of time. Yet we choose those modes to calculate
which would characterize the structure of the quasinormal spectrum
at high overtones.

\begin{figure}
\begin{center}
\includegraphics{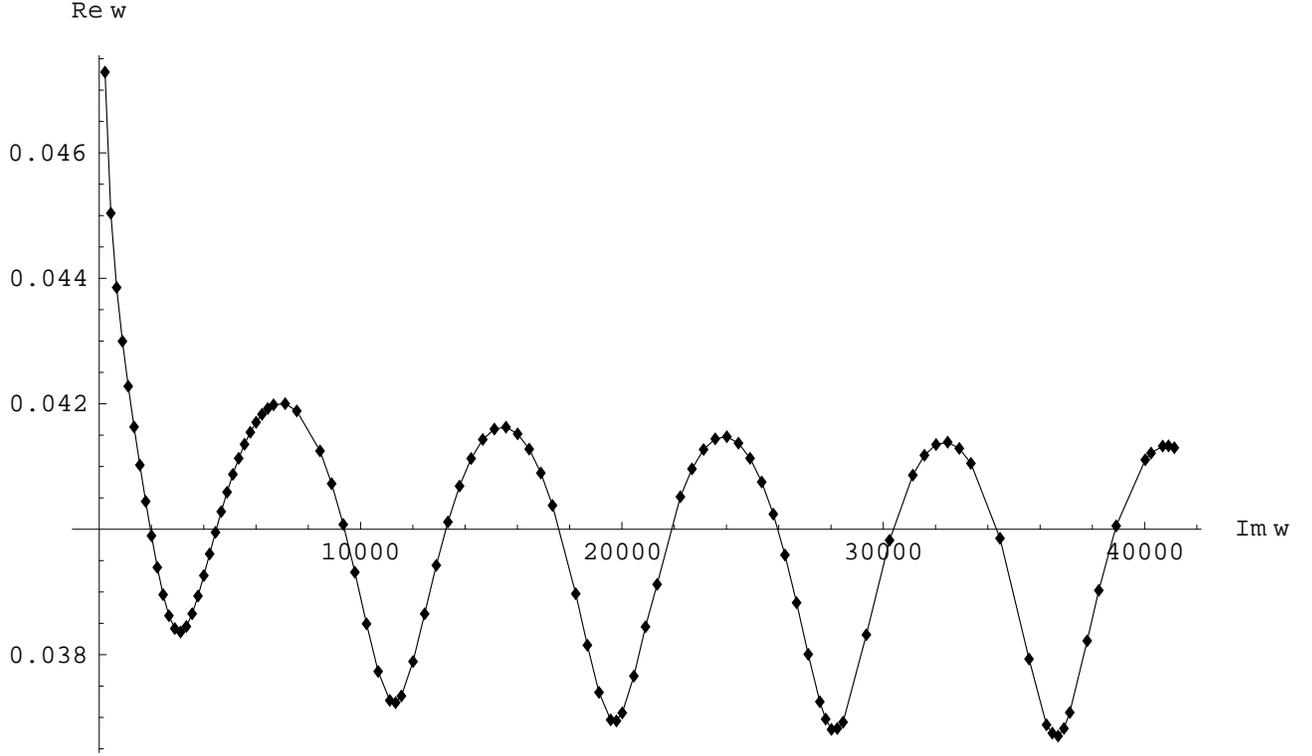}
\caption{Real part of $\omega$ as a function of imaginary part for
some values of $n=10^3, 2 \cdot 10^3, ..$. In addition to this
``large-scale'' oscillation there is another sub-oscillations when
considering nearby overtones (see for example Fig. 2). ($l=2$
gravitational modes, $\Lambda =0.02$)} \label{prd6fig2}
\end{center}
\end{figure}

\textbf{1.Gravitational perturbations.} We state that
gravitational quasinormal spectrum of SdS black hole at
asymptotically high overtones shows oscillatory behavior: {\it the
real part of $\omega$ oscillates as a function of imaginary part
and thus does not approach any constant value}. We have checked
this for very high overtones (see for example Fig. \ref{prd6fig2}
where computations performed up to $n \sim 165000$ ). The same
behavior for the real part of $\omega$ was observed in the near
extremal regime of $\Lambda$ \cite{yoshida}.


The imaginary part of $\omega$ is roughly proportional to $n$ at
large $n$ and thereby can be approximately described by the
formula:

\be\label{eg2} \Im\om  \approx - k_{e} \left(n+\frac 1
2\right)\imo, \quad n, \rightarrow \infty. \ee where $k_{e}$ is
the surface gravity at the event horizon.

Yet, this formula is not exact even for asymptotically high
overtones since the spacing between nearby overtones
$\Im{\omega_{n+1}}-\Im{\omega_{n}}$  shows very peculiar
dependence on $n$ (see as an example Fig. \textref{longfig}{4}).
From the first sight at figures 2 and 3 one could conclude that we
have with numerical noise, yet if drawing the difference
$\Im{\omega_{n+1}}-\Im{\omega_{n}}$ as a function of $n$ in a
greater scale (Fig. \textref{longfig}{4}), we see that
$\Im{\omega_{n+1}}-\Im{\omega_{n}}$ shows quite ordered repeated
structures thus the behavior is \emph{periodic}. It is important
that these periodic weaves do not have tendency of damping, i.e.
the spacing does not approach $k_{e}$ however the average value of
spacing over sufficiently large number of modes equals $k_{e}$:
\begin{equation}\label{average}
\sum_{n=N_1}^{n=N_2}\frac{(\Im{\omega_{n+1}}-\Im{\omega_{n}})}{N_2-N_1}
\approx k_{e}, \quad n ~\mathrm{is~large}.
\end{equation}
That is why the approximate formula (\ref{eg2}) is valid. The true
asymptotic formula, apparently, must have the form: \be\label{eg3}
\Im\om  \approx -\imo \left(\left(n+\frac 1 2\right) k_{e} +
f(n)\right), \quad n \rightarrow \infty. \ee where $f(n)$ is some
analytically unknown part consisting of small deviations from
$k_{e}$ similar to those shown on the Fig. \textref{longfig}{4}.
The average value of $f(n)$ in the sense of the formula
(\ref{average}) equals zero.

\begin{figure}\label{prd6fig4}
\begin{center}
\includegraphics{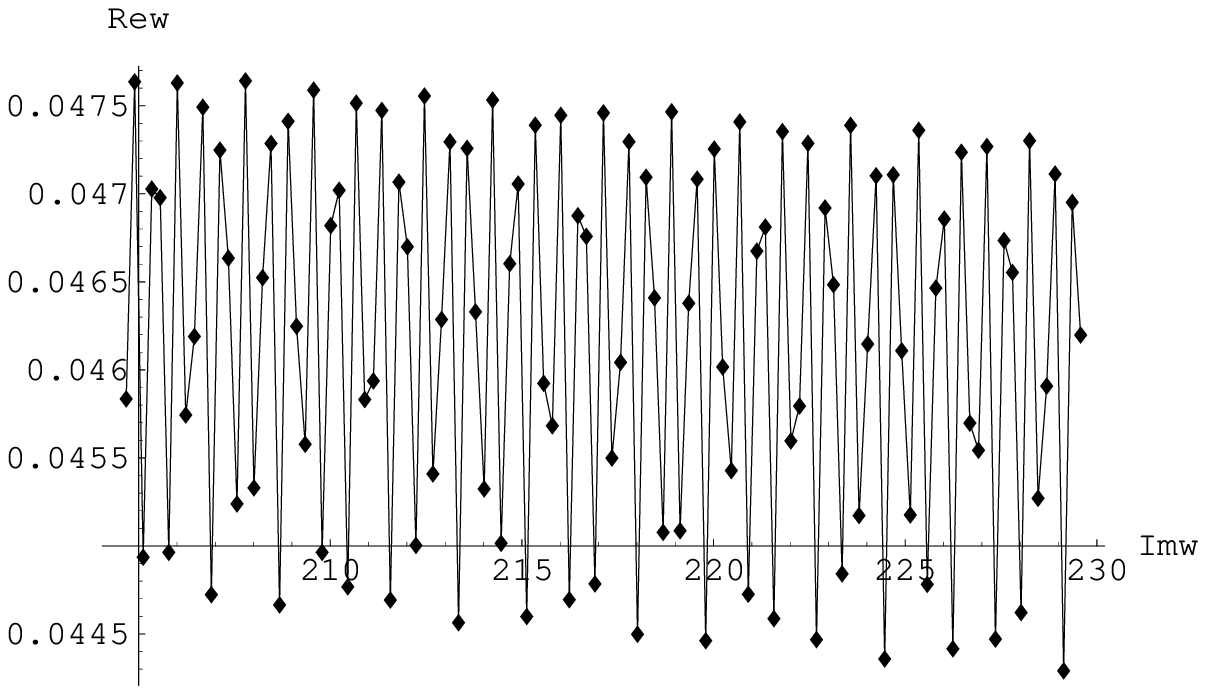}
\caption{Real part of $\omega$ as a function of imaginary part
($l=2$ gravitational modes, $\Lambda =0.02$). From the first sight
these modes can be misunderstood as just a numerical noise. Yet if
pictured in a wider scale like on Fig. 4 they show \emph{strict
periodic structure}.}
\end{center}
\end{figure}

The oscillations of real part of $\omega$ as a function of
imaginary part have very complicated form: the nearby overtones
suffer from violate oscillations with lots of maximums and
minimums similar to those shown on Fig. \ref{prd6fig4}. These
peaks form a larger wavy line like that shown on Fig.
\ref{prd6fig2}. Thus there is little hope to find a simple
analytical expression which could describe these oscillations.

An important question is how we can judge whether the computed
overtones are high enough to reflect the true asymptotic
behavior?. We are sure that at a sufficiently large $n$ the
oscillations have a stable ``profile'' which just repeats itself
with further increase of $n$. We can see the approaching of such a
``final'' profile on Fig \ref{prd6fig2}.

\textbf{2.Electromagnetic perturbations.}

Electromagnetic perturbations show rather different behavior at
high overtones: also there are oscillations of real part as a
function of imaginary part, these oscillations damp when
increasing the overtone number (see Fig. \textref{longfig}{4} as
an example). Thus the real part of $\omega $ asymptotically
approaches zero. We have checked already that at $n \sim 5000$
real part of $\omega $ is vanishing and we have not found QN modes
with non-vanishing real part at higher $n$ at all.

The imaginary part of electromagnetic  QN modes  shows the same
behavior at high overtones as the gravitational modes do. That is,
even though the imaginary part of $\omega$ is roughly proportional
to $n$ at large $n$, the spacing weaves as a function of $n$ with
average value
$\sum_{n=N_1}^{n=N_2}\frac{(\Im{\omega_{n+1}}-\Im{\omega_{n}})}{N_2-N_1}
\approx k_e, (n ~\mathrm{is~large})$ over sufficiently large
quantity of modes. On contrary to gravitational perturbations,
these weaves of the spacing of $\Im\omega$ damp and for very  high
$n$ it is approaching the equidistant spectrum:

\be \Im\om  = - k_{e} \left(n+\frac 1 2\right)\imo, \quad n
\rightarrow \infty. \ee

\begin{figure}
\begin{center}
\includegraphics{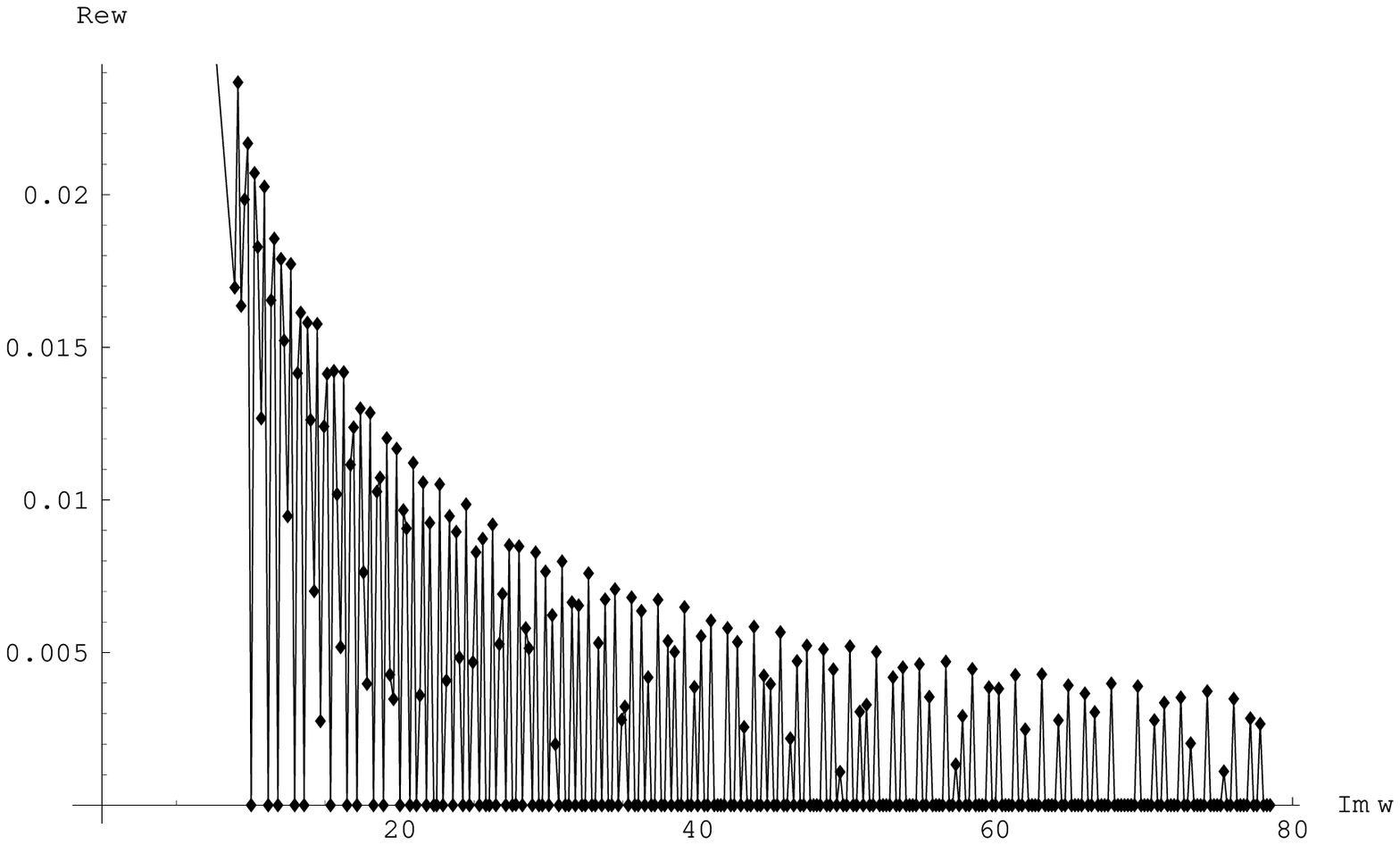}
\caption{Real part of $\omega$ as a function of imaginary part ($l=1$ electromagnetic modes, $\Lambda
=0.02$).}
\label{rrr}
\end{center}
\end{figure}

\section{Comparison with analytic formulas}

After the first version of this work has been appeared
\cite{konoplya-zhidenko}, V.Cardoso, J.Natarrio and R.Schiappa
\cite{Cardoso-Natarrio-Schiappa} managed to find an analytical
expression for $n \rightarrow \infty$ QN behavior. They found that
the gravitational modes must obey the following algebraic equation
\be\label{grav-Cardoso} \cosh \left(\frac{\pi
\omega}{k_{e}}-\frac{\pi \omega}{k_{c}}\right) + 3 \cosh
\left(\frac{\pi \omega}{k_{e}}+\frac{\pi \omega}{k_{c}}\right)=0
\ee takes place for the asymptotically high ($n \rightarrow
\infty$) gravitational modes ($k_{c}$ is the surface gravity at
the cosmological horizon.).  While for electromagnetic modes they
obtained the two possibilities \be\label{electro-Cardoso} \omega =
i (n +1/2) k_{e} \quad or \quad \omega = i (n +1/2) k_{c}. \ee
First of all the numerical data shown on Fig. \ref{prd6fig2} are
in a good agreement with the above asymptotic formula for
gravitational perturbations and the more $n$ the closer numerical
values to its analytical values, and, thereby, the better the QN
modes obey the algebraic equations  \ref{grav-Cardoso}. The
difference is quite understood, since complete coincidence is only
asymptotical. Yet the formula \ref{grav-Cardoso} does not predict
the asymptotic behavior of imaginary part as a function of
overtone number. It only connects the real and imaginary parts
within one algebraic equation. From the approximate relation
\ref{average}, one could expect that at some very high $n$ one
could observe the equdistant spectrum like $\omega = i (n+1/2)
k_{e}$. We have not observed any signs of it, since the weaves of
the spacing of imaginary part do not show any tendency to damping.

As to electromagnetic perturbations we observed that the
asymptotic formula $\omega = i (n+1/2) k_{e}$ is really true. That
is, the greater $n$ the closer QN modes to the values $\omega = i
(n+1/2) k_{e}$. This is difficult to see for not very high
overtones, but the coincidence of numerical and analytical results
is very accurate for sufficiently high overtones. An instance of
this approaching asymptotic regime is demonstrated on the
following table.

\begin{minipage}[c]{.9\textwidth}
\centerline{Comparison of numerical results shown on fig. \ref{prd6fig2}}
\centerline{with the algebraic equation of \ref{grav-Cardoso}}
\begin{longtable}{l|r}
Numerical result & Asymptotical formula\\
\hline
$0.039314- 9783.906062\imo$&$0.037864~- 9783.907040\imo$\\
$0.038492-10228.624422\imo$&$0.037008-10228.625519\imo$\\
$0.037073-20012.425574\imo$&$0.036309-20012.426688\imo$\\
$0.037659-20457.144537\imo$&$0.037042-20457.145567\imo$\\
$0.038318-29351.507086\imo$&$0.037857-29351.507866\imo$\\
$0.039824-30240.943303\imo$&$0.039382-30240.943943\imo$\\
$0.039026-38245.869248\imo$&$0.038636-38245.869869\imo$\\
$0.040054-38912.946162\imo$&$0.039657-38912.946703\imo$\\
$0.041107-40024.740368\imo$&$0.040668-40024.740824\imo$\\
$0.041214-40247.099158\imo$&$0.040764-40247.099602\imo$\\
$0.041326-40691.816716\imo$&$0.040852-40691.817141\imo$\\
$0.041329-40914.175491\imo$&$0.040843-40914.175908\imo$\\
$0.041299-41136.534267\imo$&$0.040800-41136.534676\imo$
\end{longtable}
\end{minipage}

\begin{minipage}[c]{.9\textwidth}
\centerline{Comparison of numerical results for electromagnetic
$l=1$ modes}
\centerline{with the algebraic equation of \ref{electro-Cardoso}}
\begin{longtable}{l|r}
Numerical result & Asymptotical formula\\
\hline
$\Lambda=0.02$\\
\hline
analytical & numerical \\
$0 - 2223.590486 \imo (n=10000)$&$ 0 - 2223.590586 \imo$\\
$0 - 4447.180971 \imo (n=20000)$&$ 0 - 4447.181020 \imo$\\
$0 - 44471.809715 \imo (n=200000)$&$ 0 - 44471.809720 \imo$\\
\hline
$\Lambda=0.05$\\
\hline
$0 - 1761.123825 \imo (n=10000)$&$ 0 - 1761.123919 \imo$\\
$0 - 3522.247649 \imo (n=20000)$&$ 0 - 3522.247699 \imo$\\
$0 - 35222.476494 \imo (n=200000)$&$ 0 - 35222.476499 \imo$\\
\hline
\end{longtable}
\end{minipage}

At the same time we have not found any QN modes close to  $\omega
= i (n+1/2) k_{c}$ at high $n$. This agrees with alternative
choise of \emph{either} $\omega = i (n+1/2) k_{e}$ \emph{or}
$\omega = i (n+1/2) k_{c}$ predictions.

Thus our numerical data confirm the analytic results very well.
This agreement between numerical and analytical results also
stated in the paper \cite{Cardoso-Natarrio-Schiappa}.

\section{Conclusion}
We have shown that even at very large overtone number the real
part of the gravitational quasinormal frequency of SdS black hole
oscillates as a function of imaginary part and these oscillations
do not have any tendency to damping, i.e. the real part of
$\omega$ asymptotically does not approach any constant value. On
contrary to gravitational modes, real part of electromagnetic
modes oscillates as well but these oscillations are damping with
the growing of the overtone number and the real part
asymptotically approaches zero.

In  the case of Schwarzschild black hole the interpretation of the
asymptotic value for quasinormal frequency is known \cite{Dreyer}:
the real part of it coincides with the Barbero-Immirizi parameter.
We see that such a direct correspondence should be impossible for
Schwarzschild de Sitter background, since the asymptotic value for
quasinormal frequency is not a constant. Thus the possible
connection of the QN frequency with the Barbero-Immirizi parameter
in LQG  is still an open question for Schwarzschild de Sitter
black hole.

\section*{Acknowledgements}
We would like to thank Shijun Yoshida and Karlucio Castello-Branco
for useful discussions. \mbox{R. K.} acknowledges hospitality of
the Centre for Astrophysics at Instituto Superior T\'{e}cnico in
Lisbon.

\newpage

\section{Appendix: lower overtones for electromagnetic and gravitational perturbations}

\begin{minipage}[c]{.9\textwidth}
\centerline{QNMs (electromagnetic)}
\begin{longtable}{r|rr}
\hline
$n$ & $\om M$ ($\L M^2 = 0.02$, $l = 1$) & $\om M$ ($\L M^2 = 0.02$, $l = 2$)\\
\hline
$~0$&$0.22594346-0.08410383\imo$&$0.41502231-0.08614382\imo$\\
$~1$&$0.19922380-0.26307549\imo$&$0.39900686-0.26200974\imo$\\
$~2$&$0.16431511-0.46681587\imo$&$0.37040634-0.44852516\imo$\\
$~3$&$0.13802472-0.68534232\imo$&$0.33705175-0.64953554\imo$\\
$~4$&$0.11984831-0.90808056\imo$&$0.30647368-0.86242326\imo$\\
$~5$&$0.10660509-1.13166067\imo$&$0.28135476-1.08205413\imo$\\
$~6$&$0.09637285-1.35538999\imo$&$0.26123915-1.30484284\imo$\\
$~7$&$0.08827708-1.57881240\imo$&$0.24497392-1.52897807\imo$\\
$~8$&$0.08124026-1.80213441\imo$&$0.23156338-1.75362819\imo$\\
$~9$&$0.07574049-2.02543170\imo$&$0.22028242-1.97841759\imo$\\
$10$&$0.07054630-2.24795338\imo$&$0.21062074-2.20317743\imo$\\
\hline
\hline
$n$ & $\om M$ ($\L M^2 = 0.04$, $l = 1$) & $\om M$ ($\L M^2 = 0.04$, $l = 2$)\\
\hline
$~0$&$0.20061096-0.07472608\imo$&$0.36722808-0.07623876\imo$\\
$~1$&$0.18121373-0.23012834\imo$&$0.35602130-0.23064606\imo$\\
$~2$&$0.15204473-0.40381212\imo$&$0.33469636-0.39145112\imo$\\
$~3$&$0.12839123-0.59146416\imo$&$0.30755559-0.56281639\imo$\\
$~4$&$0.11182397-0.78398818\imo$&$0.28092911-0.74463194\imo$\\
$~5$&$0.09989826-0.97695141\imo$&$0.25829220-0.93318015\imo$\\
$~6$&$0.09013754-1.17106475\imo$&$0.23991905-1.12512639\imo$\\
$~7$&$0.08368248-1.36405979\imo$&$0.22500213-1.31861131\imo$\\
$~8$&$0.07599793-1.55718153\imo$&$0.21269507-1.51274409\imo$\\
$~9$&$0.07240721-1.75138650\imo$&$0.20235705-1.70710444\imo$\\
$10$&$0.06703704-1.94219038\imo$&$0.19350620-1.90149978\imo$\\
\hline
\hline
$n$ & $\om M$ ($\L M^2 = 0.06$, $l = 1$) & $\om M$ ($\L M^2 = 0.06$, $l = 2$)\\
\hline
$~0$&$0.17089050-0.06380912\imo$&$0.31181526-0.06477818\imo$\\
$~1$&$0.15891839-0.19374672\imo$&$0.30497404-0.19515060\imo$\\
$~2$&$0.13694710-0.33439045\imo$&$0.29124711-0.32841879\imo$\\
$~3$&$0.11659604-0.48768053\imo$&$0.27176423-0.46763105\imo$\\
$~4$&$0.10180201-0.64613954\imo$&$0.25033424-0.61479838\imo$\\
$~5$&$0.09084565-0.80599862\imo$&$0.23083347-0.76843456\imo$\\
$~6$&$0.08218828-0.96605417\imo$&$0.21454829-0.92579233\imo$\\
$~7$&$0.07497191-1.12581591\imo$&$0.20119758-1.08497949\imo$\\
$~8$&$0.06886092-1.28497681\imo$&$0.19016038-1.24500614\imo$\\
$~9$&$0.06398420-1.44344850\imo$&$0.18088812-1.40538843\imo$\\
$10$&$0.06059324-1.60167622\imo$&$0.17296716-1.56589008\imo$\\
\hline
\end{longtable}
\end{minipage}
\bigskip\newline
\begin{minipage}[c]{.9\textwidth}
\begin{longtable}{r|rr}
\hline
$n$ & $\om M$ ($\L M^2 = 0.09$, $l = 1$) & $\om M$ ($\L M^2 = 0.09$, $l = 2$)\\
\hline
$~0$&$0.11053646-0.04153633\imo$&$0.20085020-0.04180306\imo$\\
$~1$&$0.10759439-0.12478386\imo$&$0.19907499-0.12548885\imo$\\
$~2$&$0.10125456-0.20875361\imo$&$0.19544199-0.20944161\imo$\\
$~3$&$0.09045512-0.29524991\imo$&$0.18978198-0.29395362\imo$\\
$~4$&$0.07890420-0.39018591\imo$&$0.18187205-0.37958230\imo$\\
$~5$&$0.07225731-0.48374011\imo$&$0.17175342-0.46745082\imo$\\
$~6$&$0.06380747-0.58344049\imo$&$0.16090315-0.55881594\imo$\\
$~7$&$0.06207673-0.67726878\imo$&$0.15106524-0.65264576\imo$\\
$~8$&$0.05339164-0.77726919\imo$&$0.14249130-0.74847390\imo$\\
$~9$&$0.05559083-0.87187144\imo$&$0.13546731-0.84457777\imo$\\
$10$&$0.04543452-0.96887722\imo$&$0.12914704-0.94185889\imo$\\
\hline
\hline
$n$ & $\om M$ ($\L M^2 = 0.11$, $l = 1$) & $\om M$ ($\L M^2 = 0.11$, $l = 2$)\\
\hline
$~0$&$0.02545378-0.00961749\imo$&$0.04614431-0.00962073\imo$\\
$~1$&$0.02542082-0.02885255\imo$&$0.04612347-0.02886223\imo$\\
$~2$&$0.02535465-0.04808787\imo$&$0.04608172-0.04810386\imo$\\
$~3$&$0.02525473-0.06732363\imo$&$0.04601892-0.06734573\imo$\\
$~4$&$0.02512026-0.08656006\imo$&$0.04593488-0.08658793\imo$\\
$~5$&$0.02495008-0.10579737\imo$&$0.04582932-0.10583057\imo$\\
$~6$&$0.02474268-0.12503586\imo$&$0.04570190-0.12507377\imo$\\
$~7$&$0.02449607-0.14427589\imo$&$0.04555216-0.14431764\imo$\\
$~8$&$0.02420768-0.16351788\imo$&$0.04537960-0.16356233\imo$\\
$~9$&$0.02387424-0.18276244\imo$&$0.04518356-0.18280800\imo$\\
$10$&$0.02349147-0.20201036\imo$&$0.04496330-0.20205483\imo$\\
\hline
\end{longtable}
\end{minipage}
\bigskip\newline
\begin{minipage}[c]{.9\textwidth}
\centerline{QNMs (gravitational)}
\begin{longtable}{r|rr}
\hline
$n$ & $\om M$ ($\L M^2 = 0.02$, $l = 2$) & $\om M$ ($\L M^2 = 0.02$, $l = 3$)\\
\hline
$~0$&$0.33839143-0.08175645\imo$&$0.54311488-0.08449572\imo$\\
$~1$&$0.31875867-0.24919663\imo$&$0.53074425-0.25536311\imo$\\
$~2$&$0.28273218-0.42948412\imo$&$0.50701532-0.43205884\imo$\\
$~3$&$0.24054151-0.62819192\imo$&$0.47483507-0.61839452\imo$\\
$~4$&$0.20194822-0.84100966\imo$&$0.43911708-0.81624020\imo$\\
$~5$&$0.16891621-1.06095994\imo$&$0.40465171-1.02429658\imo$\\
$~6$&$0.13944328-1.28417153\imo$&$0.37401137-1.23949553\imo$\\
$~7$&$0.11044868-1.50935542\imo$&$0.34769654-1.45900264\imo$\\
$~8$&$0.07677592-1.73728467\imo$&$0.32524173-1.68092084\imo$\\
$~9$&$0.00000000-1.98403566\imo$&$0.30595356-1.90413890\imo$\\
$10$&$0.05371227-2.27038762\imo$&$0.28919474-2.12803551\imo$\\
\hline
\end{longtable}
\end{minipage}
\bigskip\newline
\begin{minipage}[c]{.9\textwidth}
\begin{longtable}{r|rr}
\hline
$n$ & $\om M$ ($\L M^2 = 0.04$, $l = 2$) & $\om M$ ($\L M^2 = 0.04$, $l = 3$)\\
\hline
$~0$&$0.29889472-0.07329668\imo$&$0.48005752-0.07514635\imo$\\
$~1$&$0.28584094-0.22172415\imo$&$0.47165827-0.22639484\imo$\\
$~2$&$0.25999193-0.37709218\imo$&$0.45501064-0.38077311\imo$\\
$~3$&$0.22627597-0.54558261\imo$&$0.43107569-0.54098604\imo$\\
$~4$&$0.19342155-0.72690541\imo$&$0.40254385-0.70946230\imo$\\
$~5$&$0.16520278-0.91558017\imo$&$0.37329553-0.88660299\imo$\\
$~6$&$0.14087547-1.10757733\imo$&$0.34628879-1.07059758\imo$\\
$~7$&$0.11886442-1.30135403\imo$&$0.32264813-1.25905540\imo$\\
$~8$&$0.09700714-1.49561599\imo$&$0.30231712-1.45013447\imo$\\
$~9$&$0.07243555-1.69317330\imo$&$0.28481977-1.64268182\imo$\\
$10$&$0.03639911-1.88991383\imo$&$0.26963306-1.83602924\imo$\\
\hline
\hline
$n$ & $\om M$ ($\L M^2 = 0.06$, $l = 2$) & $\om M$ ($\L M^2 = 0.06$, $l = 3$)\\
\hline
$~0$&$0.25328922-0.06304253\imo$&$0.40717516-0.06413956\imo$\\
$~1$&$0.24574200-0.18979104\imo$&$0.40217056-0.19280739\imo$\\
$~2$&$0.23007644-0.31915725\imo$&$0.39205277-0.32276933\imo$\\
$~3$&$0.20669673-0.45518321\imo$&$0.37678925-0.45532936\imo$\\
$~4$&$0.18068220-0.60121128\imo$&$0.35698877-0.59230260\imo$\\
$~5$&$0.15723866-0.75477156\imo$&$0.33460484-0.73527072\imo$\\
$~6$&$0.13729735-0.91211670\imo$&$0.31232419-0.88423367\imo$\\
$~7$&$0.11998413-1.07114416\imo$&$0.29197277-1.03775443\imo$\\
$~8$&$0.10425546-1.23089238\imo$&$0.27412339-1.19420006\imo$\\
$~9$&$0.08916814-1.39091982\imo$&$0.25865044-1.35237729\imo$\\
$10$&$0.07372448-1.55104255\imo$&$0.24520634-1.51154481\imo$\\
\hline
\hline
$n$ & $\om M$ ($\L M^2 = 0.09$, $l = 2$) & $\om M$ ($\L M^2 = 0.09$, $l = 3$)\\
\hline
$~0$&$0.16261045-0.04136653\imo$&$0.26184253-0.04164389\imo$\\
$~1$&$0.16078859-0.12415216\imo$&$0.26057158-0.12496881\imo$\\
$~2$&$0.15704232-0.20711724\imo$&$0.25799755-0.20841194\imo$\\
$~3$&$0.15114075-0.29047481\imo$&$0.25405408-0.29207761\imo$\\
$~4$&$0.14267788-0.37469082\imo$&$0.24863818-0.37611940\imo$\\
$~5$&$0.13118467-0.46102913\imo$&$0.24161615-0.46078805\imo$\\
$~6$&$0.11824166-0.55218893\imo$&$0.23287254-0.54651859\imo$\\
$~7$&$0.10767572-0.64571873\imo$&$0.22251601-0.63402624\imo$\\
$~8$&$0.09695736-0.74179066\imo$&$0.21132493-0.72404510\imo$\\
$~9$&$0.09027189-0.83757283\imo$&$0.20044112-0.81643184\imo$\\
$10$&$0.08075778-0.93460601\imo$&$0.19042690-0.91065927\imo$\\
\hline
\end{longtable}
\end{minipage}
\bigskip\newline
\begin{minipage}[c]{.9\textwidth}
\begin{longtable}{r|rr}
\hline
$n$ & $\om M$ ($\L M^2 = 0.11$, $l = 2$)& $\om M$ ($\L M^2 = 0.11$, $l = 3$)\\
\hline
$~0$&$0.03726995-0.00961565\imo$&$0.06009145-0.00961888\imo$\\
$~1$&$0.03724934-0.02884698\imo$&$0.06007662-0.02885667\imo$\\
$~2$&$0.03720806-0.04807839\imo$&$0.06004694-0.04809452\imo$\\
$~3$&$0.03714597-0.06730994\imo$&$0.06000235-0.06733248\imo$\\
$~4$&$0.03706293-0.08654169\imo$&$0.05994279-0.08657059\imo$\\
$~5$&$0.03695863-0.10577369\imo$&$0.05986815-0.10580890\imo$\\
$~6$&$0.03683275-0.12500602\imo$&$0.05977830-0.12504745\imo$\\
$~7$&$0.03668489-0.14423876\imo$&$0.05967310-0.14428631\imo$\\
$~8$&$0.03651456-0.16347198\imo$&$0.05955236-0.16352551\imo$\\
$~9$&$0.03632114-0.18270578\imo$&$0.05941586-0.18276513\imo$\\
$10$&$0.03610395-0.20194028\imo$&$0.05926336-0.20200521\imo$\\
\hline
\end{longtable}
\end{minipage}
\newpage
\begin{minipage}{18cm}
\vspace{-4cm}\hspace{-2cm}\begin{minipage}{9cm}
\rotate{\includegraphics{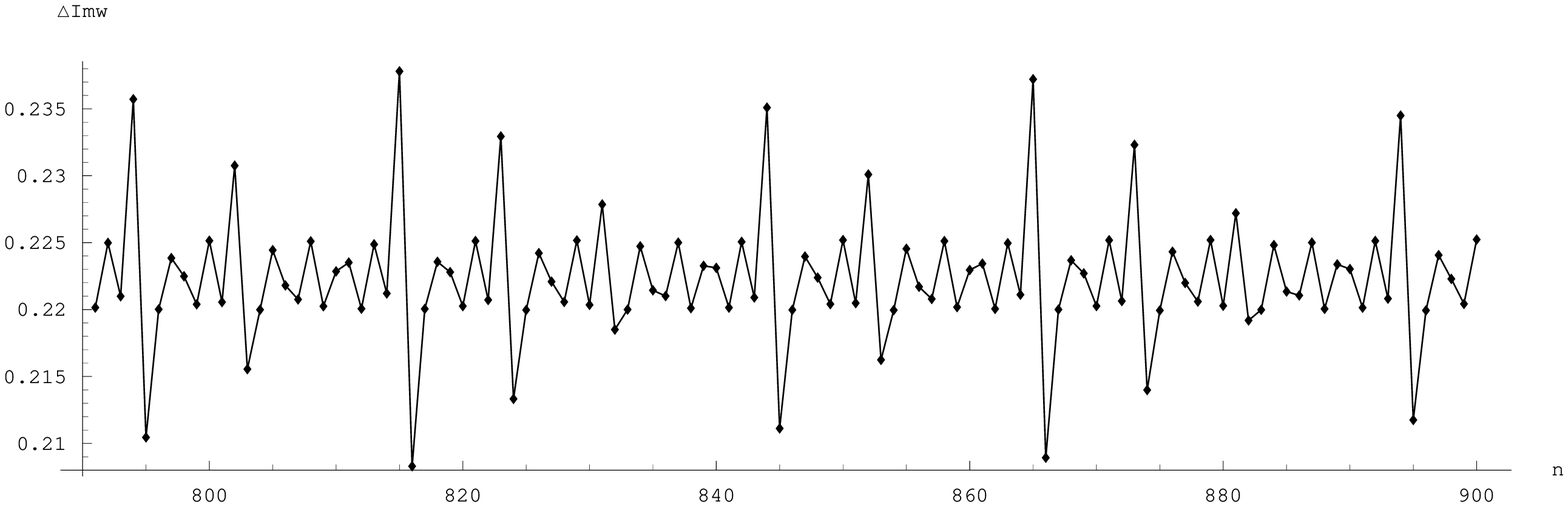}}
\end{minipage}
\begin{minipage}{8cm}
\begin{picture}(0,0)
\put(130,420){\oval(50,100)}
\put(130,690){\oval(50,100)}
\put(130,295){\oval(80,150)}
\put(130,560){\oval(80,160)}
\end{picture}
\rotate{\includegraphics{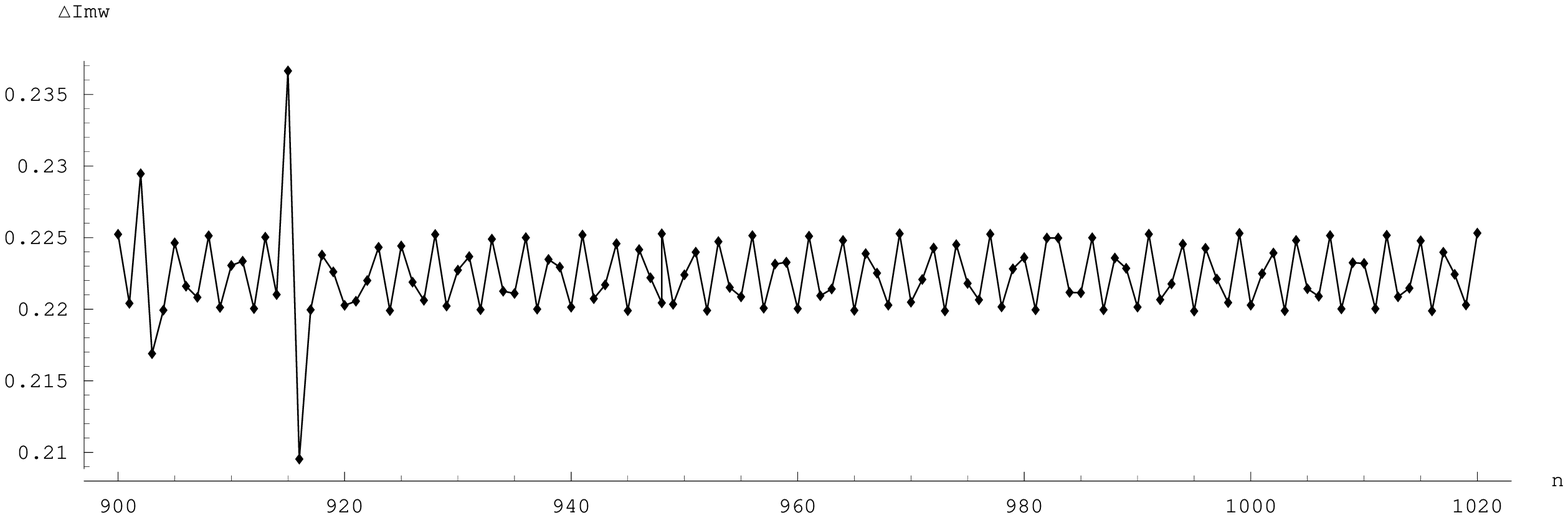}}
\end{minipage}
\end{minipage}
\begin{minipage}{16cm}
\vspace{-4cm}\hspace{-2cm}\begin{minipage}{9cm}
\rotate{\includegraphics{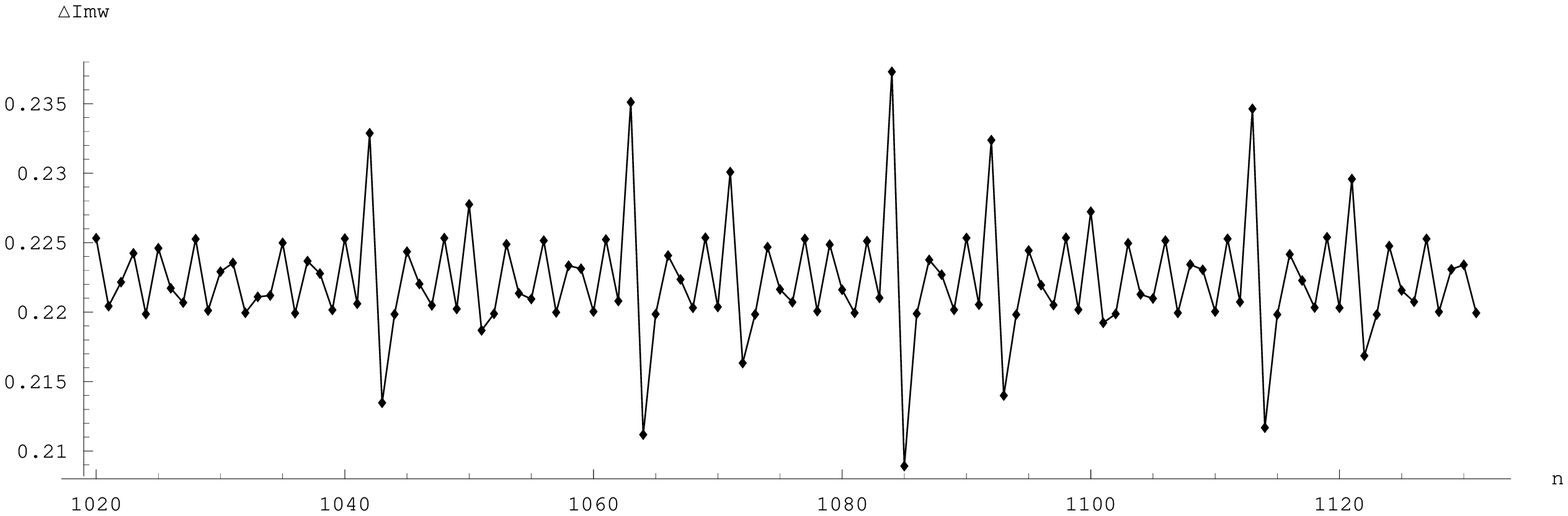}}
\end{minipage}
\begin{minipage}{8cm}\label{longfig}
\qquad\rotate{Figure 4. The spacing
$\Im{\omega_{n+1}}-\Im{\omega_{n}}$ as a function of $n$ for
gravitational perturbations, ($\Lambda=0.02$, $M=1$, $l=2$) for
large $n$.} \rotate{The spacing shows peculiar behavior which is
NOT a numerical noise, since it is strictly periodic.}
\end{minipage}
\end{minipage}

\end{document}